# The Light Hadron Mass Spectrum with Non-Perturbatively O(a) Improved Wilson Fermions

M. Göckeler[1,2], R. Horsley[3], H. Perlt[4], P. Rakow[5],
G. Schierholz[5,6], A. Schiller[4] and P. Stephenson[5]

[1] Institut für Theoretische Physik, J. W. Goethe-Universität,
D-60054 Frankfurt, Germany

[2] Institut für Theoretische Physik, RWTH Aachen,
D-52056 Aachen, Germany

[3] Institut für Physik, Humboldt-Universität,
D-10115 Berlin, Germany

[4] Institut für Theoretische Physik, Universität Leipzig,
D-04109 Leipzig, Germany

[5] Deutsches Elektronen-Synchrotron DESY,
Institut für Hochenergiephysik and HLRZ,
D-15735 Zeuthen, Germany

[6] Deutsches Elektronen-Synchrotron DESY,
D-22603 Hamburg, Germany

**Abstract**

We compute the light hadron mass spectrum in quenched lattice QCD at $\beta = 6.0$ using the Sheikholeslami-Wohlert fermionic action. The calculation is done for several choices of the coefficient $c_{SW}$, including $c_{SW} = 0$ and the recently proposed optimal value $c_{SW} = 1.769$. We find that the individual masses change by up to 30% under $O(a)$ improvement. The spectrum calculation suggests $c_{SW} \approx 1.4$ for the optimal value of the coefficient.

# 1  Introduction

Wilson fermions have some advantages over staggered fermions. For example, one is not restricted to multiples of four flavors, the flavor symmetry is exact so that one does not have to test for its restoration [1], and the flavor and Dirac structures are not intertwined on the lattice making it easier to construct composite fermionic operators. Their disadvantage is that they are accompanied by cut-off effects of $O(a)$, whereas the gauge field action and staggered fermions introduce corrections of $O(a^2)$ only.

To be able to keep the size of the lattice at a reasonable level and yet explore large physical volumes, one is forced to do simulations at moderate values of the coupling, for example at $\beta \equiv 6/g^2 = 6.0$ in the quenched theory. At this value the cut-off lies at $\approx 0.1$ fm (taking the string tension as the physical scale), which is to be compared with the typical size of a hadron of 1 fm. The result is that one has to reckon with $O(a)$ effects causing systematic errors on the level of 20% or more.

By adding the local counterterm

$$\frac{\mathrm{i}}{2} \kappa\, g\, c_{SW}(g)\, a \sum_x \bar{\psi}(x) \sigma_{\mu\nu} F_{\mu\nu}(x) \psi(x) \tag{1}$$

(we assume $r = 1$ throughout the paper) to the Wilson fermionic action, cut-off effects in on-shell quantities [2] can be reduced from $O(a)$ to $O(a^2)$ [3]. Such quantities are, for example, hadron masses and energies as well as hadronic matrix elements of local operators. In the case of hadronic matrix elements the operators have to be improved as well.

To achieve this, one has to compute the coefficient $c_{SW}(g)$ to all orders in perturbation theory, that means non-perturbatively in practice. We call the fermionic action non-perturbatively $O(a)$ improved if $c_{SW}$ is determined non-perturbatively, and we call the resulting value of $c_{SW}$ an optimal value. Perturbatively $c_{SW}$ is known to $O(g^2)$ [4]:

$$c_{SW}(g) = 1 + 0.2659\, g^2. \tag{2}$$

The lowest order perturbative result, $c_{SW} = 1$, reduces cut-off effects to $O(a/\ln a)$ only, the second order result to $O(a/\ln^2 a)$, and so on. A crude approximation of $c_{SW}$ might be obtained by reorganizing the perturbative expansion such that the leading tadpole contributions are treated non-perturbatively [5], while the remainder is computed in low-order renormalized perturbation theory. This leads to

$$c_{SW}(g) = u_0^{-3}(1 + 0.0159\, g^{*\,2}), \tag{3}$$

where $u_0 = \langle \frac{1}{3}\mathrm{Tr} U_\Box \rangle^{\frac{1}{4}}$ and $g^*$ is the coupling constant renormalized at some physical scale. Taking $g^{*\,2} = g^2 u_0^{-4}$, we obtain at $\beta = 6.0$ the value $c_{SW} = 1.518$ for $u_0 = 0.8778$.

Recently, $c_{SW}$ has been determined non-perturbatively in the quenched theory [6]. As a criterion for $O(a)$ improvement these authors use the PCAC relation, which connects the divergence of the axial vector current with the quark mass and the pseudoscalar density, and demand that it be valid exactly so that every other on-shell quantity has possible corrections of $O(a^2)$ only. The result is [7]

$$c_{SW}(g) = \frac{1 - 0.656\,g^2 - 0.152\,g^4 - 0.054\,g^6}{1 - 0.922\,g^2}\,,\ g^2 \leq 1. \tag{4}$$

At small $g$ this formula approaches the perturbative result (2). At $\beta = 6.0$ it gives $c_{SW} = 1.769$, a value which is significantly higher than the tadpole improved perturbative result. There are also other possibilities to determine $c_{SW}$ which we will address in a separate publication.

The improvement program is a systematic expansion in powers of $a$, and one has to bear in mind that it is no longer useful if $O(a^2)$ corrections are of the same magnitude as $O(a)$ effects, which will occur as $g$ increases.

In this paper we shall look at the effect of improvement on the hadron mass spectrum [1]. Our first objective is to see how sensitive masses and mass ratios are to the values of $c_{SW}$. Secondly, we would like to investigate whether a determination of $c_{SW}$ from the mass spectrum is possible.

## 2 The Calculation

This work is the first step in the aim to apply $O(a)$ improvement to our structure function calculations [9].

We work at $\beta = 6.0$ and use lattice sizes $16^3 32$ and $24^3 32$. In most of our runs we generated $100 - 200$ configurations. For the pure Wilson case we have accumulated $O(5000)$ configurations at $\kappa = 0.155, 0.153$ and $0.1515$ on the $16^3 32$ lattice in our attempt to determine the gluon distribution functions of the nucleon [10]. For the gauge field update we use a combination of 16 overrelaxation sweeps followed by a three-hit Metropolis update. This procedure is repeated 50 times to generate a new configuration.

The implementation of the improvement term (1) in our calculation is described in ref. [11]. For the matrix inversion we mainly used the minimal residue algorithm, except for the lightest quark mass on the larger lattice where we used the BiCGstab algorithm [12]. At $c_{SW} = 3$, our largest $c_{SW}$ value, we experienced convergence problems with the minimal residue algorithm. So we experimented with the BiCGstab inversion algorithm, but this did not seem to improve the situation significantly.

We used Jacobi smearing for source and sink with 50 iterations and a smearing parameter $\kappa_s = 0.21$ [13], giving a radius of the hadrons of about four lattice spacings. We employed

---
[1] For a similar investigation at $\beta = 5.7$, which reached us after this paper was completed, see ref. [8].

| | | | $c_{SW} = 0$ | | | | |
|---|---|---|---|---|---|---|---|
| $V$ | $\kappa$ | $m_\pi$ | $m_\rho$ | $m_N$ | $m_{a_0}$ | $m_{a_1}$ | $m_{b_1}$ |
| $16^3 32$ | 0.1487 | 0.637(2) | 0.682(2) | 1.061(9) | 0.891(24) | 0.940(21) | 0.943(32) |
| | 0.1515 | 0.5033(4) | 0.5682(7) | 0.902(2) | 0.817(7) | 0.851(7) | 0.849(13) |
| | 0.1530 | 0.4221(4) | 0.5058(8) | 0.798(2) | 0.763(11) | 0.797(6) | 0.809(7) |
| | 0.1550 | 0.2966(5) | 0.4227(15) | 0.652(3) | 0.735(15) | 0.717(12) | 0.736(9) |
| $24^3 32$ | 0.1550 | 0.292(2) | 0.418(5) | 0.638(8) | 0.610(48) | 0.657(33) | 0.659(35) |
| | 0.1558 | 0.229(2) | 0.384(7) | 0.555(12) | 0.616(90) | 0.613(41) | 0.638(38) |
| | 0.1563 | 0.179(3) | 0.358(11) | 0.488(22) | 0.88(15) | 0.584(52) | 0.615(44) |
| c. l. | 0.15717(3) | 0 | 0.328(5) | 0.421(15) | *0.656(19)* | *0.622(15)* | *0.646(14)* |

Table 1: The hadron masses in lattice units for pure Wilson fermions with $c_{SW} = 0$. In the bottom row we give $\kappa_c$ and the mass values extrapolated to the chiral limit. The numbers in roman (*italic*) are from three-parameter (two-parameter) fits.

| | | | $c_{SW} = 1.769$ | | | | |
|---|---|---|---|---|---|---|---|
| $V$ | $\kappa$ | $m_\pi$ | $m_\rho$ | $m_N$ | $m_{a_0}$ | $m_{a_1}$ | $m_{b_1}$ |
| $16^3 32$ | 0.1300 | 0.701(4) | 0.782(5) | 1.17(2) | | | |
| | 0.1310 | 0.624(4) | 0.715(5) | 1.062(17) | | | |
| | 0.1320 | 0.545(5) | 0.644(8) | 0.974(16) | | | |
| | 0.1324 | 0.501(2) | 0.613(4) | 0.906(13) | 0.785(17) | 0.817(12) | 0.831(23) |
| | 0.1333 | 0.410(3) | 0.547(5) | 0.791(16) | 0.74(3) | 0.750(15) | 0.778(17) |
| | 0.1342 | 0.299(3) | 0.488(10) | 0.674(27) | 0.85(8) | 0.683(20) | 0.744(26) |
| $24^3 32$ | 0.1342 | 0.302(2) | 0.492(5) | 0.695(9) | 0.82(3) | 0.715(19) | 0.758(16) |
| | 0.1346 | 0.238(2) | 0.470(9) | 0.642(16) | 1.00(8) | 0.684(26) | 0.745(20) |
| | 0.1348 | 0.188(7) | 0.457(20) | 0.605(22) | 1.52(20) | 0.664(34) | 0.736(29) |
| c. l. | 0.13529(3) | 0 | 0.429(11) | 0.548(25) | *0.845(38)* | *0.629(19)* | *0.716(18)* |

Table 2: The hadron masses in lattice units for non-perturbatively $O(a)$ improved Wilson fermions with $c_{SW} = 1.769$, according to eq. (4). In the bottom row we give $\kappa_c$ and the mass values extrapolated to the chiral limit. The numbers in roman (*italic*) are from three-parameter (two-parameter) fits.

| | $c_{SW} = 1.92$ | | | | | | |
|---|---|---|---|---|---|---|---|
| $V$ | $\kappa$ | $m_\pi$ | $m_\rho$ | $m_N$ | $m_{a_0}$ | $m_{a_1}$ | $m_{b_1}$ |
| $16^3 32$ | 0.1290 | 0.644(1) | 0.734(2) | 1.121(9) | 0.888(9) | 0.946(10) | 0.959(11) |
| | 0.1300 | 0.553(1) | 0.661(3) | 1.001(10) | 0.819(11) | 0.879(11) | 0.892(12) |
| | 0.1310 | 0.451(2) | 0.586(4) | 0.875(11) | 0.755(14) | 0.812(13) | 0.827(14) |
| | 0.1320 | 0.328(2) | 0.516(10) | 0.735(20) | 0.793(65) | 0.763(29) | 0.771(34) |
| c. l. | 0.13322(6) | 0 | 0.430(32) | 0.539(88) | *0.611(33)* | *0.668(27)* | *0.681(30)* |

Table 3: The hadron masses in lattice units for improved Wilson fermions with $c_{SW} = 1.92$. In the bottom row we give $\kappa_c$ and the mass values extrapolated to the chiral limit. The numbers in roman (*italic*) are from three-parameter (two-parameter) fits.

| | $c_{SW} = 2.25$ | | | | | | |
|---|---|---|---|---|---|---|---|
| $V$ | $\kappa$ | $m_\pi$ | $m_\rho$ | $m_N$ | $m_{a_0}$ | $m_{a_1}$ | $m_{b_1}$ |
| $16^3 32$ | 0.1250 | 0.639(1) | 0.739(3) | 1.097(10) | 0.885(16) | 0.940(22) | 0.946(24) |
| | 0.1260 | 0.535(2) | 0.658(4) | 0.966(12) | 0.800(18) | 0.848(22) | 0.861(25) |
| | 0.1265 | 0.476(2) | 0.617(4) | 0.893(11) | 0.774(26) | 0.802(26) | 0.808(33) |
| | 0.1270 | 0.414(2) | 0.576(5) | 0.807(15) | 0.738(38) | 0.754(26) | 0.750(43) |
| | 0.1277 | 0.305(3) | 0.514(12) | 0.684(19) | | 0.733(61) | 0.697(33) |
| c. l. | 0.12861(5) | 0 | 0.435(32) | 0.461(80) | *0.597(56)* | *0.602(49)* | *0.597(44)* |

Table 4: The hadron masses in lattice units for improved Wilson fermions with $c_{SW} = 2.25$. In the bottom row we give $\kappa_c$ and the mass values extrapolated to the chiral limit. The numbers in roman (*italic*) are from three-parameter (two-parameter) fits.

| | $c_{SW} = 3$ | | | |
|---|---|---|---|---|
| $V$ | $\kappa$ | $m_\pi$ | $m_\rho$ | $m_N$ |
| $16^3 32$ | 0.1150 | 0.705(4) | 0.840(4) | 1.239(9) |
| | 0.1155 | 0.643(4) | 0.783(4) | 1.172(9) |
| | 0.1160 | 0.555(6) | 0.727(3) | 1.054(9) |
| | 0.1165 | 0.470(12) | 0.684(6) | 0.936(13) |
| | 0.1170 | 0.408(7) | 0.603(7) | 0.858(12) |
| | 0.1173 | 0.342(12) | 0.569(13) | 0.758(16) |
| c. l. | 0.11826(13) | 0 | *0.464(13)* | *0.558(25)* |

Table 5: The hadron masses in lattice units for improved Wilson fermions with $c_{SW} = 3$. In the bottom row we give $\kappa_c$ and the mass values extrapolated to the chiral limit. The latter values are from two-parameter fits.

| $c_{SW}$ | $\kappa_c$ | | |
|---|---|---|---|
| | eq. (5) | eq. (6) | eq. (7) |
| 0 | 0.15693(1) | 0.15734(5) | 0.15717(3) |
| 1.769 | 0.13522(1) | 0.13534(4) | 0.13529(3) |
| 1.92 | 0.13306(2) | 0.13333(11) | 0.13322(6) |
| 2.25 | 0.12850(2) | 0.12866(8) | 0.12861(5) |
| 3 | 0.11799(4) | 0.11862(39) | 0.11826(13) |

Table 6: The critical values of $\kappa$, $\kappa_c$, for the linear (eq. (5)), chiral (eq. (6)) and phenomenological fit (eq. (7)) for our various $c_{SW}$ parameters.

both relativistic and non-relativistic wave functions [9, 13], except for the high statistics runs where we only looked at the non-relativistic wave function in order to save computer time.

We calculated the masses of $\pi$, $\rho$, nucleon, $a_0$, $a_1$ and $b_1$. In the following we shall mainly focus on the $\pi$, $\rho$ and nucleon masses. The $c_{SW}$ values we considered were 0 (pure Wilson), 1.769 (the value of eq. (4)), 1.92, 2.25 and 3. In addition, we shall make use of the UKQCD results [14] for the $\pi$ and $\rho$ mass at $c_{SW} = 1.4785$, which is their estimate of the tadpole improved perturbative value.

## 3 Results

We present our results in tables 1 – 5. The errors on the data are bootstrap errors. For the meson masses we found very little difference between using relativistic and non-relativistic wave functions, and we settled for relativistic wave functions (except for the high statistics runs). For the nucleon we have chosen non-relativistic wave functions which performed slightly better.

To obtain the critical value of $\kappa$, $\kappa_c$, and the hadron masses in the chiral limit, we extrapolate our data to zero $\pi$ mass. We expect

$$m_\pi^2 = b \left( \frac{1}{\kappa} - \frac{1}{\kappa_c} \right). \tag{5}$$

Using this relation gives a rather poor fit of the data, and we can see that there is slight curvature in a plot of $m_\pi^2$ against $1/\kappa$. Quenched chiral perturbation theory predicts [15]

$$m_\pi^2 = b' \left( \frac{1}{\kappa} - \frac{1}{\kappa_c} \right)^{\frac{1}{1+\delta}}, \tag{6}$$

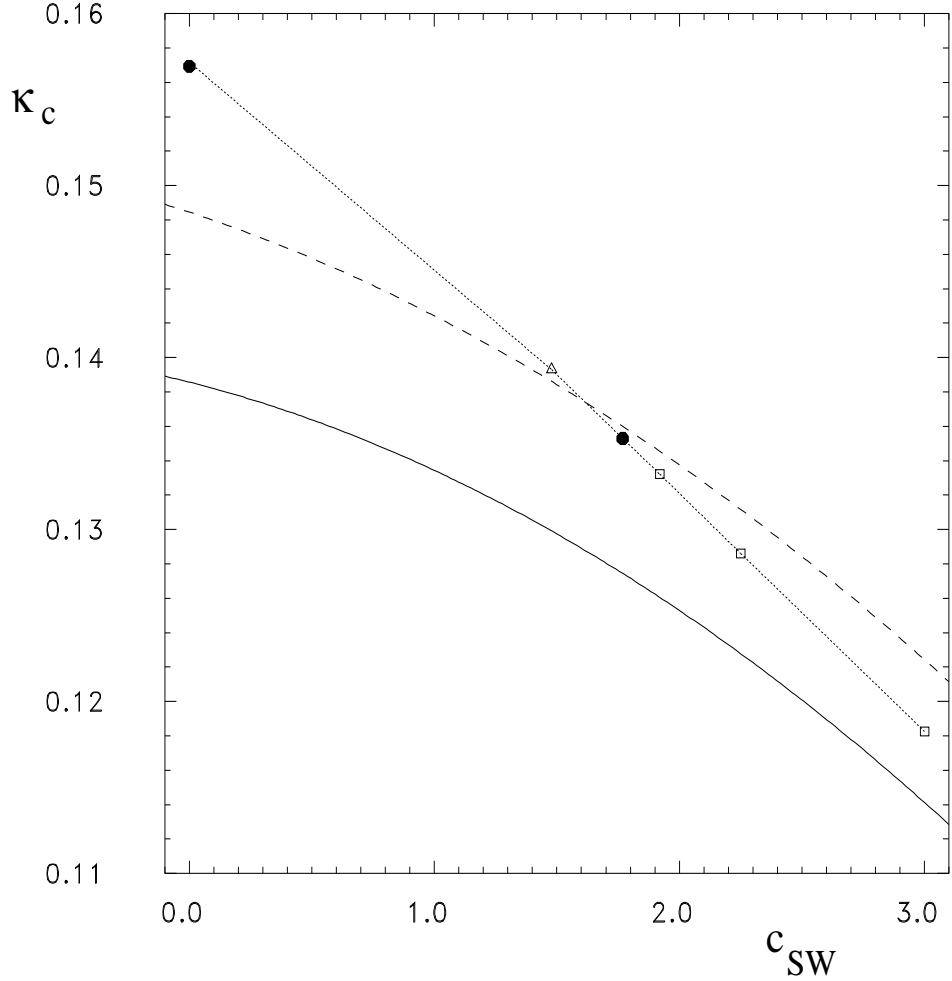

Figure 1: The critical value of $\kappa$ against $c_{SW}$. The points at $c_{SW} = 0$ and 1.769 ($\bullet$) include data at small $\pi$ masses on the $24^3 32$ lattice, the points at 1.92, 2.25 and 3 ($\square$) are from the $16^3 32$ lattice. The data point at $c_{SW} = 1.4785$ is from UKQCD [14]. The solid curve is the prediction of perturbation theory, the dashed curve that of tadpole improved perturbation theory. The dotted line connecting the data points is meant to guide the eye.

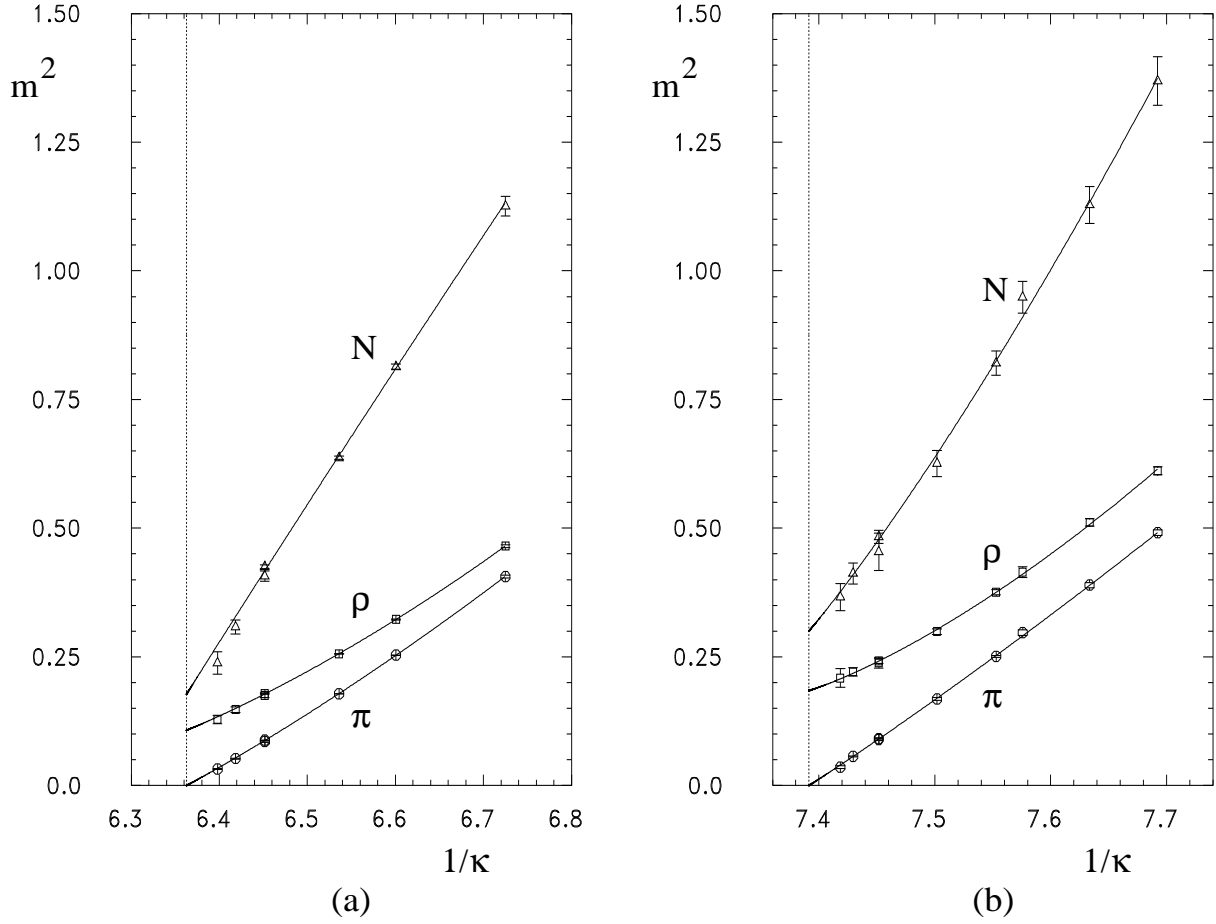

Figure 2: Fits and chiral extrapolations of the hadron masses for (a) $c_{SW} = 0$ and (b) $c_{SW} = 1.769$.

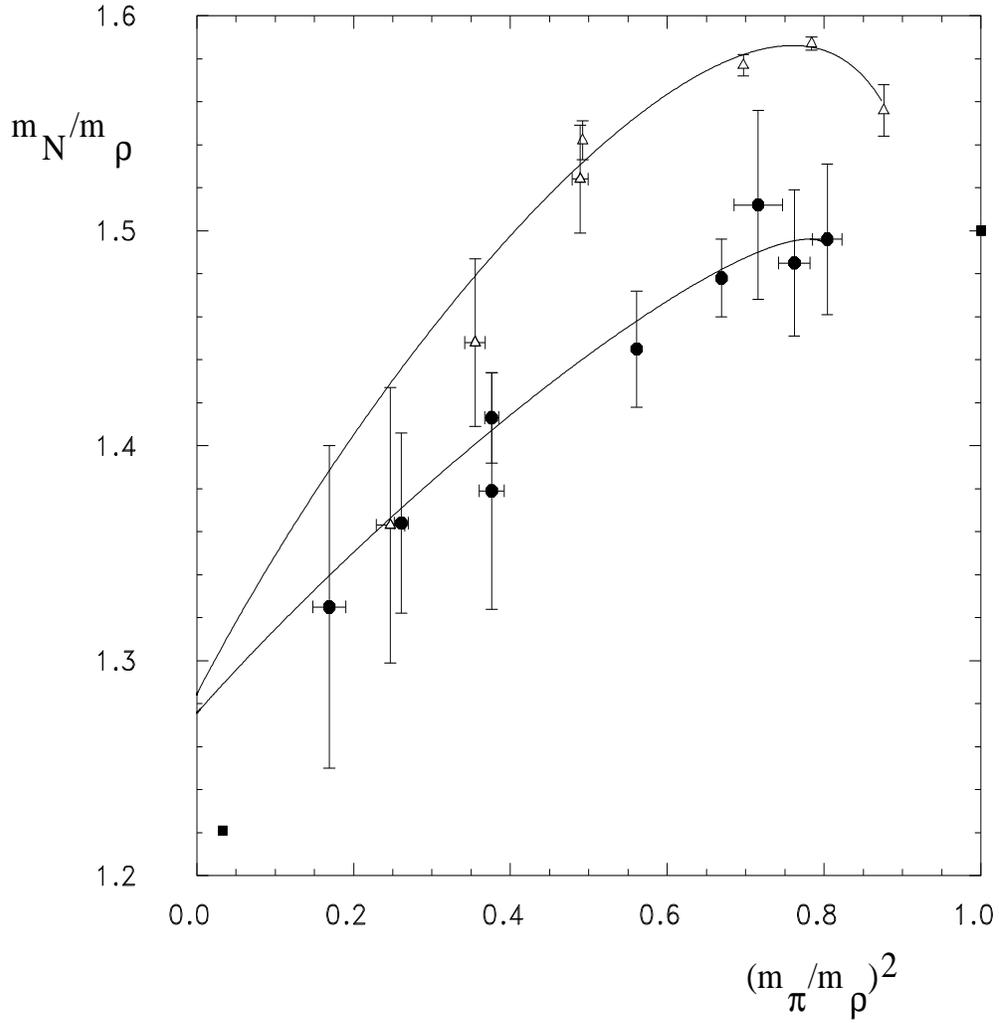

Figure 3: APE plot for $c_{SW} = 0$ ($\triangle$) and $c_{SW} = 1.769$ ($\bullet$). The errors are bootstrap errors. The solid boxes represent the experimental value and the expected result in the heavy quark limit, respectively.

where $\delta$ is small and positive. We made fits using this formula but found that $\delta$ was always negative. We concluded that our $\kappa$ values are too far from $\kappa_c$ for the formula to be applicable. This is in agreement with observations made by other authors [16]. As an alternative parameterization of the curvature we used the phenomenological fit

$$\frac{1}{\kappa} = \frac{1}{\kappa_c} + b_2 m_\pi^2 + b_3 m_\pi^3. \tag{7}$$

In table 6 we give the values of $\kappa_c$ for the different fits. The linear fits give $\chi^2/d.o.f.$ values of up to 20. The other two fits give both acceptable values of $\chi^2$, but the phenomenological fits give slightly better results. In the following we shall take $\kappa_c$ from the phenomenological fits.

In fig. 1 we plot $\kappa_c$ against $c_{SW}$. We observe a rather linear behavior of $\kappa_c$ with $c_{SW}$. It is interesting to compare the numerical values with the perturbative result

$$\kappa_c = \frac{1}{8}[1 + g^2(0.108571 - 0.028989\, c_{SW} - 0.012064\, c_{SW}^2)] \tag{8}$$

and tadpole improved perturbative result

$$u_0 \kappa_c = \frac{1}{8}[1 + g^{*\,2}(0.025238 - 0.028989\, c_{SW}\, u_0^3 - 0.012064\, (c_{SW}\, u_0^3)^2)]. \tag{9}$$

Neither formula agrees well with the data, but the tadpole improved line gives rough agreement in the neighborhood of $c_{SW} = 1.769$.

We fit the rest of the hadron masses by the formula

$$m_H^2 = m_H^{0\,2} + b_2 m_\pi^2 + b_3 m_\pi^3. \tag{10}$$

This gives a better fit to the data than the ansatz [17]

$$m_H = m_H^0 + b_2' m_\pi^2 + b_3' m_\pi^3. \tag{11}$$

Note that the two formulae differ only by terms of $O(m_\pi^4)$. For $c_{SW} = 3$ as well as for the $a_0$, $a_1$ and $b_1$ masses only a two-parameter fit with $b_3$ set to zero was reasonable. The mass values in the chiral limit are also given in tables 1 – 5. We consider our extrapolations for $c_{SW} = 0$ and 1.769 to be most reliable, because on the larger volume we could explore relatively small $\pi$ masses.

We show in fig. 2 the $\pi$, $\rho$ and nucleon masses for $c_{SW} = 0$ and 1.769 together with the fits. It is important to have many $\kappa$ values in the fit to do a reliable extrapolation. At the third smallest quark mass we have results on two different volumes. The values agree within the errors. This indicates that all our results on the $16^3 32$ lattice do not suffer from finite size effects.

In fig. 3 we show the APE plot for $c_{SW} = 0$ and 1.769 together with our chiral extrapolations. We find rather different results for the two $c_{SW}$ values. However, in the chiral limit the nucleon to $\rho$ mass ratio is about the same. The $c_{SW} = 1.769$ result appears to go more smoothly to the heavy quark limit.

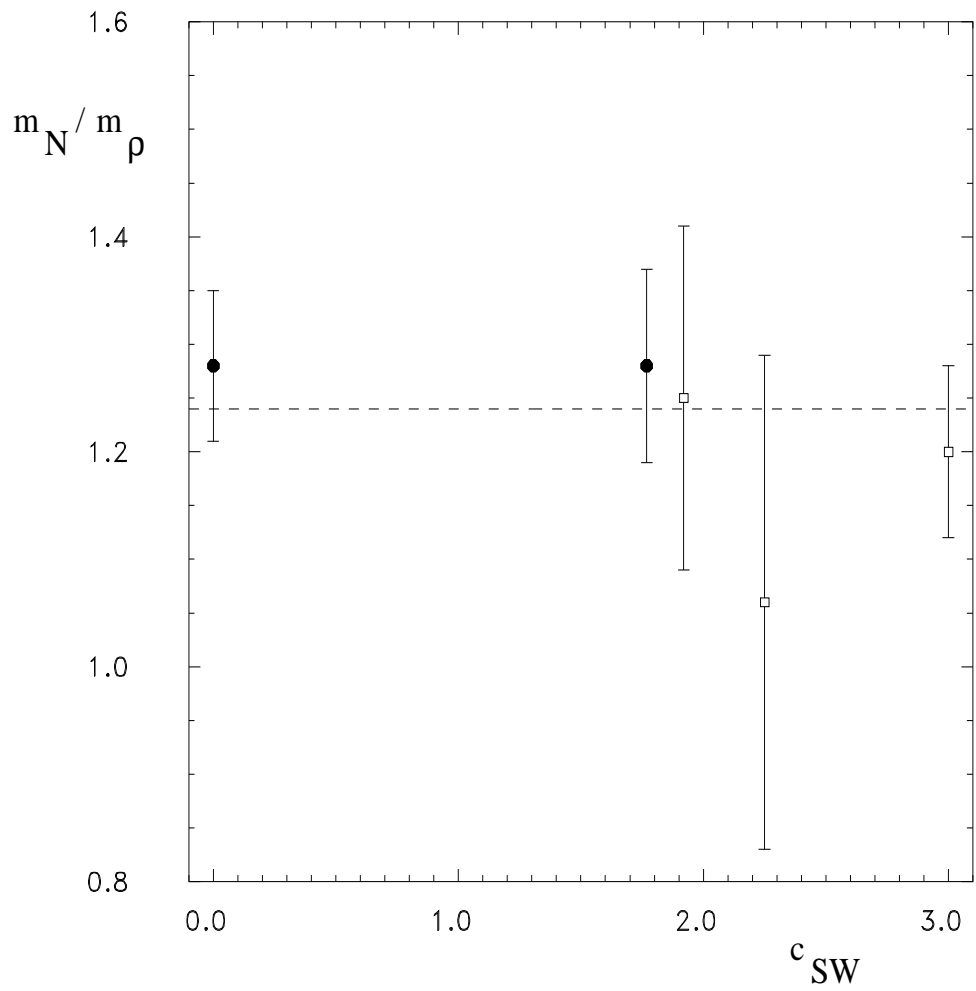

Figure 4: The nucleon to $\rho$ mass ratio in the chiral limit against $c_{SW}$. The symbols are as in fig. 1. The dashed line is the experimental value.

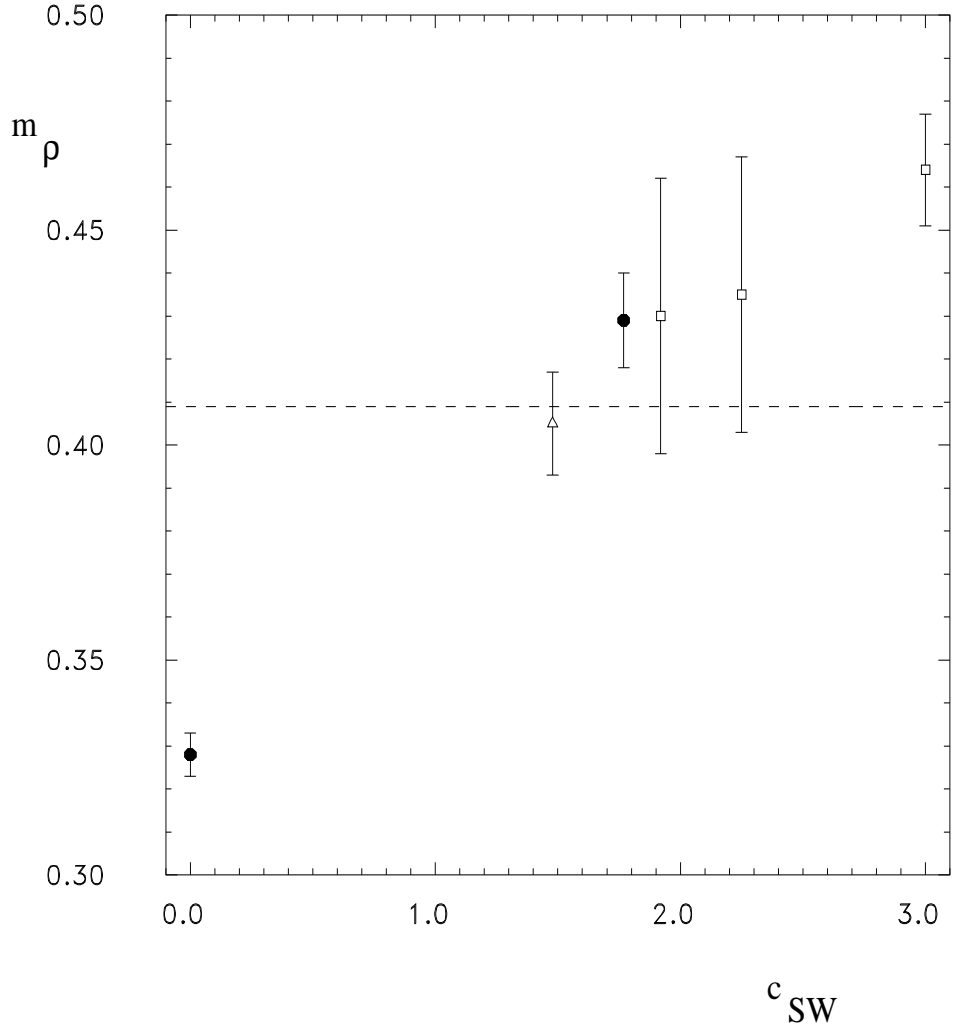

Figure 5: The $\rho$ mass in the chiral limit against $c_{SW}$. The symbols are as in fig. 1. The dashed line is the experimental value, where the lattice spacing is taken from the string tension: $K = 0.0515(28)$ [18], with $K = (427\,\mathrm{MeV})^2$ as given by the Cornell potential [19].

As we noted in the APE plot, the nucleon to $\rho$ mass ratio in the chiral limit was the same at $c_{SW} = 0$ and 1.769. In fig. 4 we show the ratio for all our $c_{SW}$ values. The results agree with the experimental value within the errors and seem to be independent of $c_{SW}$. Away from the chiral limit the mass ratio depends more strongly on $c_{SW}$. This means that if one is only interested in mass ratios in the chiral limit, then Wilson fermions seem to give reasonable extrapolations.

Although the nucleon to $\rho$ mass ratio is independent of $c_{SW}$, when we look at the $\rho$ (and nucleon) mass itself in the chiral limit we see a strong $c_{SW}$ dependence, as is shown in fig. 5. Varying $c_{SW}$ from 0 to 1.769 we observe a 30% increase of the mass. Even between the tadpole improved perturbative result $c_{SW} \approx 1.5$ and $c_{SW} = 1.769$ the mass changes by about 10%. Expressing the $\rho$ mass in terms of the string tension $K$, which has discretization errors of $O(a^2)$ only, gives the dashed line. If one assumes that the effect of quenching can be neglected, one can use these results to estimate the optimal value of $c_{SW}$ up to possible $O(a^2)$ corrections.

## 4 Discussion

We have done a systematic investigation of the $c_{SW}$ dependence of the light hadron mass spectrum. We see that at $\beta = 6.0$ it is important to choose a good value of $c_{SW}$.

Disregarding errors due to quenching and $O(a^2)$ corrections, our best guess for the optimal value of $c_{SW}$ is $c_{SW} \approx 1.4$. This result is somewhat lower than the value given by eq. (4). If the difference is attributed to $O(a^2)$ effects, we would estimate that $O(a^2)$ corrections can be half as big as $O(a)$ corrections at $\beta = 6.0$. This would mean that the on-shell improvement program does not make much sense at much lower values of $\beta$, for example at $\beta = 5.7$.

We consider our work only to be a first step. Computing masses in the chiral limit is difficult. A lesson we have learned is that one needs many $\kappa$ values for a reliable chiral extrapolation. A more precise determination of the optimal value of $c_{SW}$ would require much higher statistics.

## Acknowledgement

This work was supported in part by the Deutsche Forschungsgemeinschaft. The numerical calculations were performed on the Quadrics computers at DESY-Zeuthen. We wish to thank the operating staff for their support. We furthermore like to thank M. Lüscher and R. Sommer for communicating to us their results for $c_{SW}(g)$ prior to publication.